\documentclass[prb,onecolumn,preprint,preprintnumbers,amsmath,amssymb,floats,superscriptaddress,notitlepage]{revtex4-1}
\usepackage{graphicx}
\usepackage{dcolumn}
\usepackage{bm}
\usepackage{times}

\begin{document}

\title{Emergent Rank-5 `Nematic' Order in URu$_2$Si$_2$}

\author{H. Ikeda$^{1,*}$, M.-T. Suzuki$^2$, R. Arita$^{3}$, T. Takimoto$^{4}$, T. Shibauchi$^{1}$, and Y. Matsuda}
\affiliation{
Department of Physics, Kyoto University, Kyoto 606-8502, Japan \\
$^{2}$CCSE, Japan Atomic Energy Agency, 5-1-5 Kashiwanoha, Kashiwa, Chiba 277-8587, Japan \\
$^{3}$Department of Applied Physics, University of Tokyo, Tokyo 113-8656, Japan \\
$^{4}$Asia Pacific Center for Theoretical Physics, POSTECH, Pohang 790-784, Korea \\
$^*${\rm To whom correspondence should be addressed. E-mail:}
{\sf\small hiroaki@scphys.kyoto-u.ac.jp}
}

\date{October 31, 2011}%

\begin{abstract}
\bf
Novel electronic states resulting from entangled spin and orbital degrees of freedom are hallmarks of strongly correlated $\boldsymbol{f}$-electron systems.  A spectacular example is the so-called `hidden-order' phase transition in the heavy-electron metal URu$_2$Si$_2$, which is characterized by the huge amount of entropy lost at $\boldsymbol{T_{HO}=17.5}$~K  \cite{rf:Palstra,rf:Maple}.  However, no evidence of magnetic/structural phase transition has been found below $\boldsymbol{T_{HO}}$ so far.  The origin of the hidden-order phase transition has been a long-standing mystery in condensed matter physics.  Here, based on a first-principles theoretical approach, we examine the complete set of multipole correlations allowed in this material.  The results uncover that the hidden-order parameter is a rank-5 multipole (dotriacontapole) order with `nematic' $\boldsymbol{E^-}$ symmetry, which exhibits staggered pseudospin moments along the [110] direction. This naturally provides  comprehensive explanations of all key features in the hidden-order phase including anisotropic magnetic excitations, nearly degenerate antiferromagnetic-ordered state, and spontaneous rotational-symmetry breaking. 
\end{abstract}

\maketitle

\vfill 

\newpage

In the rare-earth and actinide compounds, $f$-electrons behave like well-localized moments at high temperatures.  As the temperature is lowered,  $f$-electrons begin to delocalize due to the hybridization with conduction electron wavefunctions.  At yet lower temperatures the $f$-electrons become itinerant, forming a narrow conduction band with heavy effective electron mass, which is largely enhanced from the free-electron mass.   Notable many-body effects within the narrow band lead to a plethora of fascinating physical phenomena including multipole order, quantum phase transition and unconventional superconductivity.   Among them, perhaps the appearance of a `hidden-order' (HO) state in URu$_2$Si$_2$ is one of the most mysterious phenomena.  Identification of the microscopic order parameter and mechanism that derives the HO transition continue to be a central question in the strongly correlated $f$-electron systems.

There are several unique features that appear to be clues for understanding the HO in URu$_2$Si$_2$.  In the paramagnetic state above $T_{HO}$,  the magnetic susceptibility exhibits the Ising-like anisotropy  \cite{rf:Palstra,rf:Ramirez}.  In the HO state below $T_{HO}$, an electronic excitation gap is formed on a large portion of the Fermi surface (FS) \cite{rf:Schmidt,rf:Aynajian} and most of the carriers disappear \cite{rf:Schoenes,rf:Behnia,rf:Kasahara}.  Closely related to this, the gap formation also occurs in the magnetic excitation spectra  at commensurate and incommensurate wave numbers, $\boldsymbol{Q}_C$ = (0 0 1) and $\boldsymbol{Q}_{IC}$ = (0.6 0 0), respectively, as revealed by the neutron inelastic scattering \cite{rf:Broholm,rf:Bourdarot,rf:Wiebe}. The HO ground state changes to the large-moment antiferromagnetic (AFM) state with the ordering vector $\boldsymbol{Q}_C$ upon applying hydrostatic pressure \cite{rf:Amitsuka,rf:Motoyama,rf:Hassinger}, but the FS  has a striking similarity between these different phases \cite{rf:Ohkuni,rf:Hassinger2}, implying that the HO is nearly degenerate with the AFM order.  The magnetic torque measurements reveal the `nematicity', which breaks the in-plane rotational (tetragonal) symmetry in the HO \cite{rf:Okazaki}.  The challenge for the theory has been to identify the order parameter which explains all the above key features.

The theories that have been proposed to describe the HO state can be divided into two prevailing approaches; one is based on the localized 5$f$-electron model and the other the itinerant one \cite{rf:Mydosh}.   Recent angle-resolved-photoemission-spectroscopy results clearly demonstrate that all 5$f$ electrons are itinerant \cite{rf:Kawasaki} and the crystalline electric field, which is a signature of the localized nature,  has never been observed.   Moreover,  the nuclear magnetic resonance (NMR)  measurements \cite{rf:Takagi} show a formation of the coherent heavy-electron state well above $T_{HO}$.  Therefore it is natural to discuss the electronic structure based on the itinerant picture.  However, reliable calculation of the physical quantities by taking into account the complicated band structure is a difficult task.  For this purpose, we use  a state-of-the-art {\it ab initio} downfolding \cite{rf:Info} and dissect the electronic structure obtained from the density-functional theory (DFT) calculations.   The obtained tight-binding Hamiltonian is constructed from 56 orbitals of U 5$f$, U $6d$, Ru $4d$ and Si $3p$.   Introducing the on-site Coulomb interactions between 5$f$-electrons, we obtain a realistic itinerant model, i.e. 56-band Anderson-lattice model including the spin-orbit interaction.  Based on this realistic model Hamiltonian, magnetic and multipole correlations are analyzed by the random-phase approximation (RPA) and beyond. To account for the mass renormalization effect in the Fermi-liquid theory, the energy and temperature scale is reduced by a factor of 10 \cite{rf:Palstra,rf:Ohkuni,rf:Hassinger2} throughout this study, which makes comparisons to the experiments straightforward.

Figure~1 displays the paramagnetic FS and the band structure near the Fermi level, respectively. The energy bands crossing the Fermi level have mainly the total angular momentum $j=5/2$ multiplet of U $5f$.  Each $j_z$ component of $j=5/2$ multiplet is weighed by color.  It turns out that each separated FS is mainly composed of a rather specific $j_z$ component without large mixing, except for the outer FS around $Z$ point (Fig.~1).  Such a $j_z$ component map is quite useful in that we are able to capture valuable information such as which parts of the FS play an essential role for the HO formation.  Indeed, the disentanglement of FS orbital characters has also been an important theoretical advance to understand the electronic properties in iron-pnictide superconductors \cite{rf:Kuroki}. 

First we discuss the RPA analysis of rank-1 (dipole) correlation, which is the conventional static magnetic correlation.  The regime with $j_z=\pm 5/2$ shown by red in outer FS around $Z$ point is well nested with outer FS around $\Gamma$ point  by the vector $\boldsymbol{Q}_C$  as indicated by the arrow in Fig.~1 \cite{rf:Oppeneer} .  This nesting gives rise to a sharp peak of the correlation parallel to the $c$ axis  (dipole $J_z$) at $Z$ (0 0 1)  shown in Fig.~2{\bf a}.  Another salient feature is the hump structure at around (0.6 0 0) and the equivalent points, whose $\boldsymbol{Q}$-vectors coincide with $\boldsymbol{Q}_{IC}$.  We point out  that these peak and hump structures in the paramagnetic phase are directly related to the magnetic excitation gap at $\boldsymbol{Q}_C$ and $\boldsymbol{Q}_{IC}$ in the HO phase \cite{rf:Wiebe} because the gap opening occurs at the nested parts of the FS.  We also note that the in-plane magnetic correlations, $J_x$ and $J_y$,  are not enhanced in contrast to $J_z$ (Fig.~2{\bf b}), which is also consistent with the Ising-like magnetic susceptibility \cite{rf:Palstra,rf:Ramirez} and polarized-neutron measurements \cite{rf:Bourdarot}.

Next we examine the higher-rank multipole correlations.  According to the group theory, there are 36 multipole moments up to the fifth rank in $j=5/2$ subspace (Table S1).  Figures~2 {\bf c}-{\bf f} depict the correlations between the basis functions belonging to rank-2 (quadrupole), 3 (octupole), 4 (hexadecapole) and 5 (dotriacontapole), respectively.   What is remarkable is that similar to the dipole $J_z$,  the $\boldsymbol{Q}_C$ correlation at $Z$-point is strongly enhanced in some cases such as $O_{20}$ (rank-2), $H_{x(y)b}$ (rank-4), $D_4$ (rank-5), etc.   Generally, these bases are  mixed in the tetragonal symmetry, as shown by finite off-diagonal terms (red lines).   The multipole correlations obtained by the diagonalization is depicted in Fig.~2{\bf g}, in which each correlation at $\boldsymbol{Q}_C$ is classified by the irreducible representations and the dominant component is denoted in parentheses.  At low temperatures, $A_2^-(J_z)$, $E^-(D_{x(y)})$ and $A_1^-(D_4)$ symmetries exhibit the first, second and third strongest enhancement.   The former and the latter two correspond to the AFM and dotriacontapole states, respectively.  Within the RPA, the AFM state always overcomes the dotriacontapole states.  To go beyond the RPA, we take into account the mode-mode coupling by including the Maki-Thompson type vertex corrections. For this purpose,  we calculate the maximum eigenvalue $\lambda$ of the Bethe-Salpeter equation for a staggered particle-hole pairing with use of the RPA result as the kernel.   At $\lambda=1$, a phase transition occurs from the paramagnetic  to the corresponding eigenstate. Temperature dependence of $\lambda$ of each symmetry is shown in Fig.~3{\bf a}.   As the temperature is lowered,  $\lambda$ of $E^-(D_{x(y)})$ is most strongly enhanced and the condition $\lambda=1$ is fulfilled at finite temperature, indicating a phase transition to the $E^-(D_{x(y)})$ state.   We emphasize that $E^-(D_{x(y)})$ symmetry breaks the in-plane fourfold symmetry, which naturally accounts for the `nematicity' observed in the magnetic torque results \cite{rf:Okazaki,rf:Thalmeier}.  In addition, $E^-(D_{x(y)})$ state breaks the time reversal symmetry, which is consistent with the NMR measurements \cite{rf:Takagi}. These lead us to conclude that the HO is $E^-(D_{x(y)})$ dotriacontapole order.

The present calculations also reproduce well other key features of the HO, i.e. near degeneracy of the HO and AFM states and the anisotropic temperature dependence of the uniform susceptibility.  Fig.~3{\bf a} demonstrates that  $\lambda$ of  $E^-(D_{x(y)})$ is very close to that of $A_2^-(J_z)$.  This indicates that both states are nearly degenerate and small perturbation can change the HO to AFM state.   Indeed, we can construct a phase diagram by tuning the interactions \cite{rf:Info}, which is consistent with the pressure-temperature phase diagram (Fig.~3{\bf b}).   The temperature dependence of the uniform susceptibility $\chi_c(0)$ parallel to the $c$ axis exhibits a broad maximum at around 40K, whilst $\chi_{ab}(0)$ perpendicular to the $c$ axis is smaller and nearly temperature independent (Fig.~3{\bf c}), in good agreement with experiments \cite{rf:Palstra,rf:Ramirez}.  The low-temperature decrease of $\chi_c(0)$ arises from the deep dip structure in the density of states (DOS) near the Fermi level (Fig.~S1{\bf b}).  The Ising-like susceptibility including its temperature dependence has been discussed in terms of the crystalline electric field excitations of the localized 5$f$-electrons so far.  However, the present results demonstrate that the susceptibility can be well accounted for by the itinerant scenario.

Why is such a high rank  multipole state (rank-5) with $E^-(D_{x(y)})$-symmetry realized in URu$_2$Si$_2$?   In the paramagnetic state, the FS nesting with $\boldsymbol{Q}_C$ vector plays an essential role on the multipole fluctuations.   What is crucially important is that the FS regions connected by this $\boldsymbol{Q}_C$ are dominated by the $\pm 5/2$ components, as shown in Fig.~1. In this situation, we can consider a subspace consisting of only two components $j_z=5/2$ and $-5/2$, which allows us to map $j_z=\pm 5/2$ to pseudospin $\uparrow$ and $\downarrow$. Then the dipole $J_z$ is described by the Pauli matrix $\sigma_z$, as it has only diagonal elements corresponding to $\pm 5/2 \Leftrightarrow \pm 5/2$. In contrast, $D_{x(y)}$ is given by $\sigma_{x(y)}$, representing off-diagonal components describing the $\pm 5/2 \Leftrightarrow \mp 5/2$ transition \cite{rf:Info}, which accompanies the angular momentum change of $5\hbar$ allowed only in rank-5.  In this pseudospin space, the staggered $J_z$ state corresponds to the Neel order along the $c$ axis. On the other hand, the $D_{x(y)}$ state corresponds to the in-plane order breaking the rotational symmetry, where in-plane pseudospin moments are antiferromagnetically coupled along the $c$ axis (Fig.~3{\bf d}). Thus the pressure-induced first-order transition from HO to AFM state can be explained by the pseudospin staggered moment flip from the in-plane to the out-of-plane direction.
The experimentally observed `nematicity' along the [110]-direction corresponds to the linear combination $D_{[110]}=\frac{1}{\sqrt{2}}(D_{x}+D_{y})$ of the two-fold degenerate $D_{x}$ and $D_{y}$.
The HO parameter is then represented by 
\begin{equation}
\phi_{[110]}(\bm{k})=\sum_{\alpha,\beta=\uparrow,\downarrow}\langle f_{\bm{k}\alpha}^\dagger \sigma_{[110]}^{\alpha\beta} f_{\bm{k}+\bm{Q}_c \beta}\rangle, ~~~~~~
\sigma_{[110]}=\frac{\sigma_x+\sigma_y}{\sqrt{2}},
\end{equation}
where $f_{\bm{k}\alpha}$ is an annihilation operator for an $f$-electron with momentum $\bm{k}$ and pseudospin $\alpha$.  
It should be noted that under in-plane 180$^\circ$ rotation, the pseudospins change their direction, which discriminates this state from a nematic phase in the strict sense. However, its staggered nature leads to the twofold `nematic' symmetry of the bulk susceptibility as observed experimentally.

Figure~4{\bf a} displays the FS in the HO state, which is calculated by applying the effective multipole field so as to open the gap of 4~meV observed by the scanning tunneling microscopy \cite{rf:Schmidt,rf:Aynajian}. The lattice doubling in the AFM phase with $\boldsymbol{Q}_C$ also occurs in the HO phase.   Most part of the FS having $j_z=\pm 5/2$ components disappears as a result of the gap opening at the nested parts of the paramagnetic FS.  Around the $\Gamma$-point, a small electron and a large hole ($\alpha$) pockets, the FS with a cage-like structure and four electron pockets ($\beta$) exist.  The FS in the HO phase bears a striking resemblance to that in the AFM state, consistent with the quantum oscillation measurements.  However, the broken fourfold symmetry in the HO state can be seen clearly in the FS with cage-like structure (Fig.~4{\bf b}), in sharp contrast to the AFM state \cite{rf:Info}.    

The present approach based on the first-principles calculation is able to give a comprehensive explanation to the problem of HO, which has been a quarter century mystery.   Why has the HO been hidden for a long time?  The reason appears to be  that the order parameter of the present high-rank multipole state is extremely difficult to detect directly by the conventional experimental techniques,  such as  resonant X-ray and neutron measurements.    The {\it itinerant} multipole ordering with `nematicity' revealed in the present study is a new type of electron ordering, which is expected to be ubiquitously present in the strongly correlated electron systems\cite{rf:Fradkin}  when spin and orbital degrees of freedom are entangled.

\newpage
\section*{Acknowledgements}
We thank K. Ueda, K. Haule, G. Kotliar, M. Sigrist and T.M. Rice for helpful discussions and suggestions.  This work was supported by Grant-in-Aid for the Global COE program ``The Next Generation of Physics, Spun from Universality and Emergence'', Grant-in-Aid for Scientific Research on Innovative Areas ``Heavy Electrons'' (20102002,20102006) from MEXT, and KAKENHI from JSPS.

\section*{Author contributions}
H.I. and R.A. both developed a methodology of the DFT+RPA and beyond.  M.-T.S. analyzed the fermiology in some ordered states.  T.T. provided group-theoretical arguments of multipoles.  H.I., T.S. and Y.M. wrote the text.  All authors contributed to critical discussion of the physical interpretation of the results.

\begin{figure}[ht]
\caption{Paramagnetic FS and energy band dispersion colored by weight of $j_z$ component. Red, green and blue color gauges correspond to $j_z=\pm 5/2$, $\pm 3/2$ and $\pm 1/2$ components, respectively.  The FS is constructed from two hole FSs around $Z$, and the other four electron FSs. Small (blue) electron pockets centered at $X$ and $\Gamma$ are constructed from $j_z=\pm 1/2$ component, and the inner (green) hole pocket around $Z$ is from $j_z=\pm 3/2$.  The outer hole pocket around $Z$ is a hybridized band between $\pm 3/2$ and $\pm 5/2$. The outer electron FS around $\Gamma$ is mainly composed of $j_z=\pm 5/2$, and partially hybridized with $\pm 1/2$. Two outer FSs around $\Gamma$ and $Z$ are partially nested with $\boldsymbol{Q}_C=$ (0 0 1) indicated by arrow. }

\caption{({\bf a}) Momentum dependence of magnetic susceptibility at $T=12$K for $J_z$. A complete set of multipole correlations at $T=12$K are shown along high-symmetry line for rank-1 dipole ({\bf b}), rank-2 quadrupole ({\bf c}), rank-3 octupole ({\bf d}), rank-4 hexadecapole ({\bf e}) and rank-5 dotriacontapole ({\bf f}) basis functions.  Off-diagonal correlations between different bases are also shown in red curves. ({\bf g}) Diagonalized multipole correlations as a function of temperatures, where $D_{x(y)}\equiv (D_{x(y)a1}+D_{x(y)a2}+D_{x(y)b})/\sqrt{3}$.  These all correlations have been obtained within the RPA calculations for $U=U'\simeq 2.3$ and $J=J'=0$ in units of $1/\rho_f$, where $\rho_f$ is the total $f$-electron DOS at the Fermi level. }

\caption{({\bf a}) Temperature dependence of the maximum eigenvalue $\lambda$ of the Bethe-Salpeter equation beyond the RPA including the Maki-Thompson type vertex corrections for $U=U'\simeq 2.4$ and $J=J'=0$. Note that $\lambda=1$ gives a phase transition temperature to the corresponding eigenstate. ({\bf b}) By increasing the Hund's coupling $J$ \cite{rf:Info}, the small difference of $\lambda$ between the rank-5 $E^-(D_{x(y)})$ and rank-1 $A_2^-(J_z)$ states can be reversed, which may account for the pressure-induced AFM state. ({\bf c}) Temperature dependence of uniform susceptibilities parallel and perpendicular to the $c$ axis calculated beyond the RPA. ({\bf d}) Schematic configurations of the $\pm 5/2$ pseudospin moments in the HO (left) and AFM (right) states are shown by the arrows. In both states, the pseudospins order antiferromagnetically along the $c$ axis, but the direction of the staggered moments in the pseudospin space differs between the two: along [110] for the HO state and along [001] for the AFM state (center). }

\caption{$E^-(D_{[110]})$-state FS ({\bf a}) and the band dispersion ({\bf b}) colored by weight of the $j_z$ components. ({\bf a}) The Bulilluoin zone is folded with $\boldsymbol{Q}_C=$ (0 0 1) and a two-dimensional cut of FSs is shown in the $k_z=0$ plane.  The electronic structure exhibits the in-plane four-fold symmetry breaking, which can be most easily seen in the cage FS. Two FSs around $M$ (blue and green lines in the right bottom) have almost no splitting along $XM$, which is in contrast to the large splitting found for the $A_2^-$ AFM and the $E^+$ state \cite{rf:Info}. ({\bf b}) The dispersion along $\Gamma M$ line (left panel) highlights the excitation gap of $\sim4$meV (arrow).  The enlarged figure near the cage FS (right panel) shows a pronounced anisotropic $\Gamma M$ dispersion between $\Sigma$ (red) and $\Sigma'$ (green) lines.}
\end{figure}

\begin{center}
\newpage\includegraphics[width=0.8\textwidth]{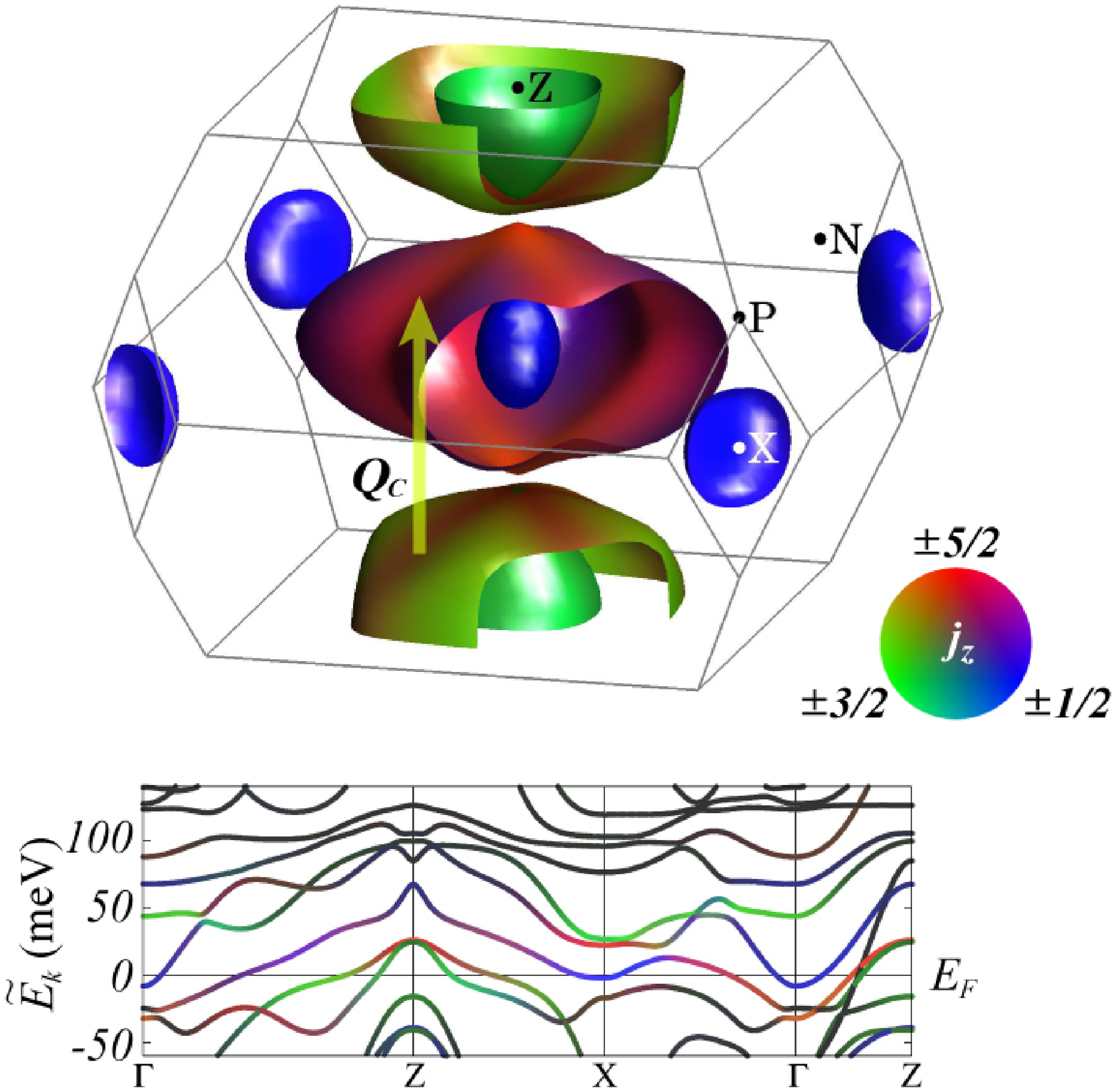}
\newpage\includegraphics[width=\textwidth]{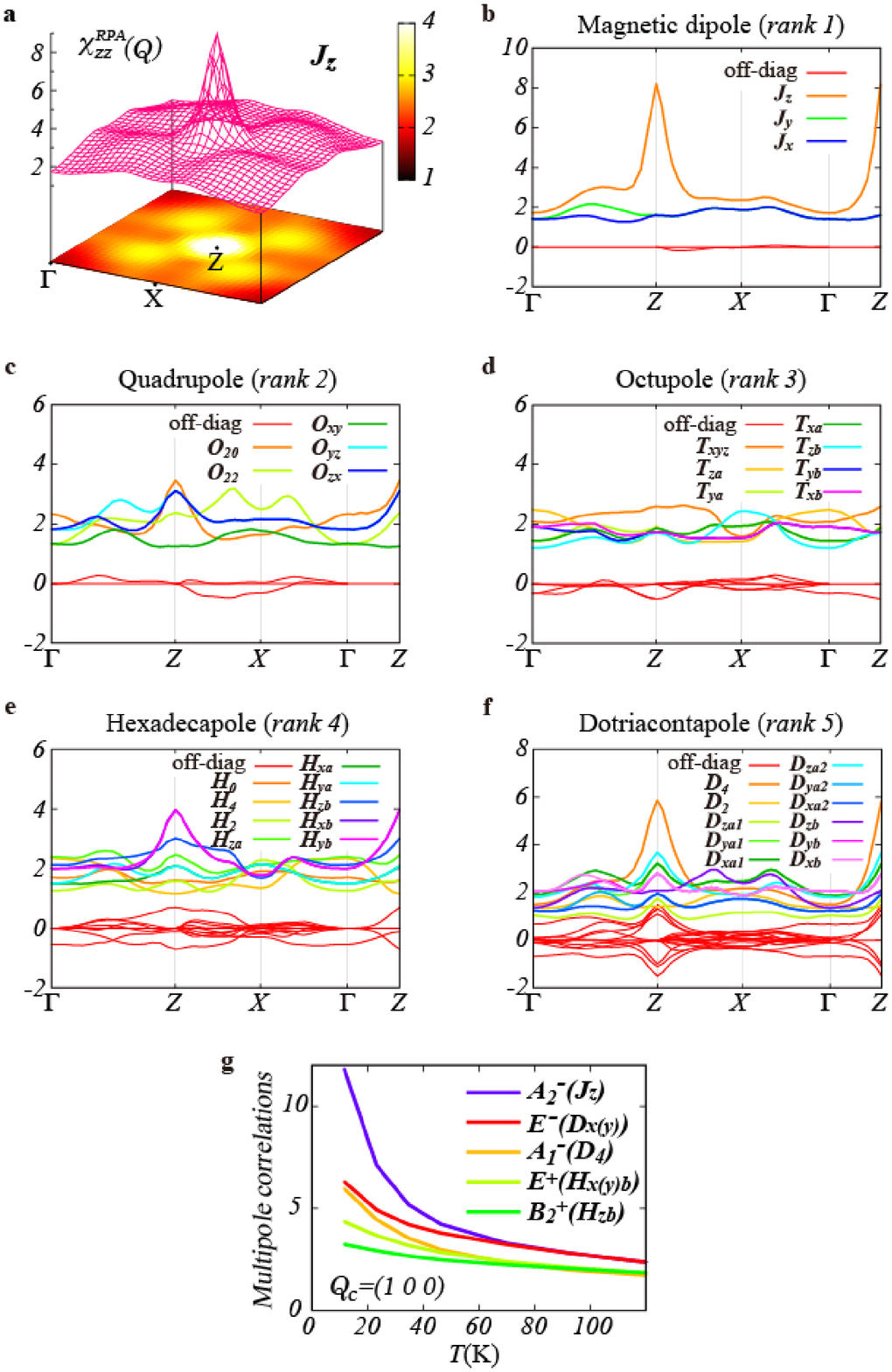}
\newpage\includegraphics[width=\textwidth]{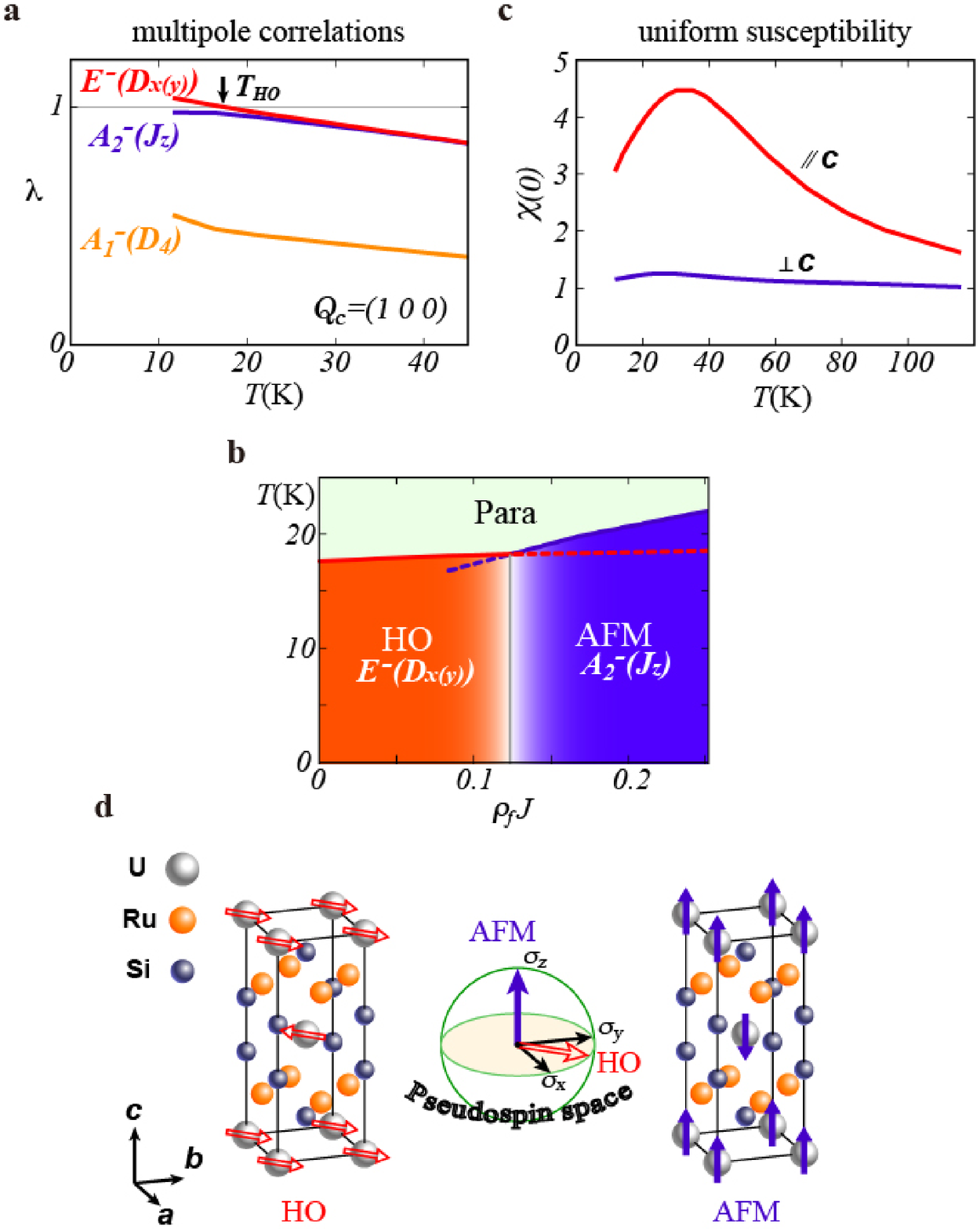}
\newpage\includegraphics[height=\textheight]{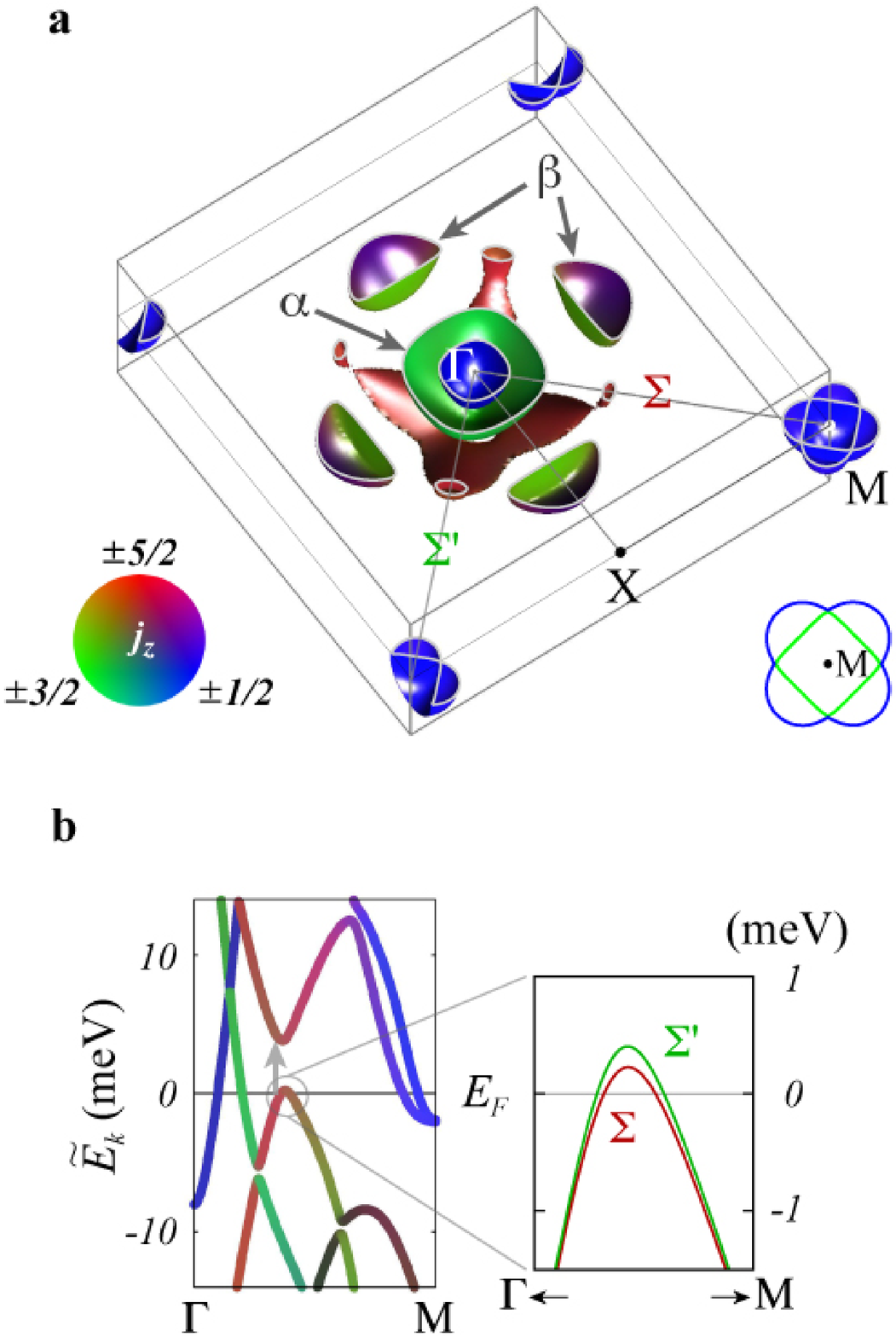}
\end{center}

\clearpage

\renewcommand{\thetable}{S\arabic{table}}
\renewcommand{\thefigure}{S\arabic{figure}}
\renewcommand{\theequation}{S\arabic{equation}}
\renewcommand{\thefootnote}{S\arabic{footnote})}
\setcounter{section}{0}
\setcounter{table}{0}
\setcounter{figure}{0}
\setcounter{equation}{0}

{\bf SUPPLEMENTARY INFORMATION} 
\begin{center}
{\bf\large Emergent Rank-5 `Nematic' Order in URu$_2$Si$_2$}
\end{center}

\section{Electronic band-structure calculation and $\boldsymbol{AB}$ $\boldsymbol{INITIO}$ downfolding}
First, we perform the {\it ab initio} band-structure calculation in the paramagnetic state of URu$_2$Si$_2$ using the {\sc wien2k} package \cite{rf:wien2k}, in which the relativistic full-potential (linearized) augmented plane-wave (FLAPW) + local orbitals method is implemented.  The crystallographical parameters are the space group No.139, $I4/mmm$, the lattice constants, $a=4.126$\AA, $c=9.568$\AA, and Si internal position, $z=0.371$ \cite{rf:Cordier}.  Red line in Fig.~S1{\bf a} depicts the electronic band structure, which is very similar to previous works \cite{rf:Oppeneer,rf:Ohkuni}.  
\begin{figure}[t]
\includegraphics[width=\textwidth]{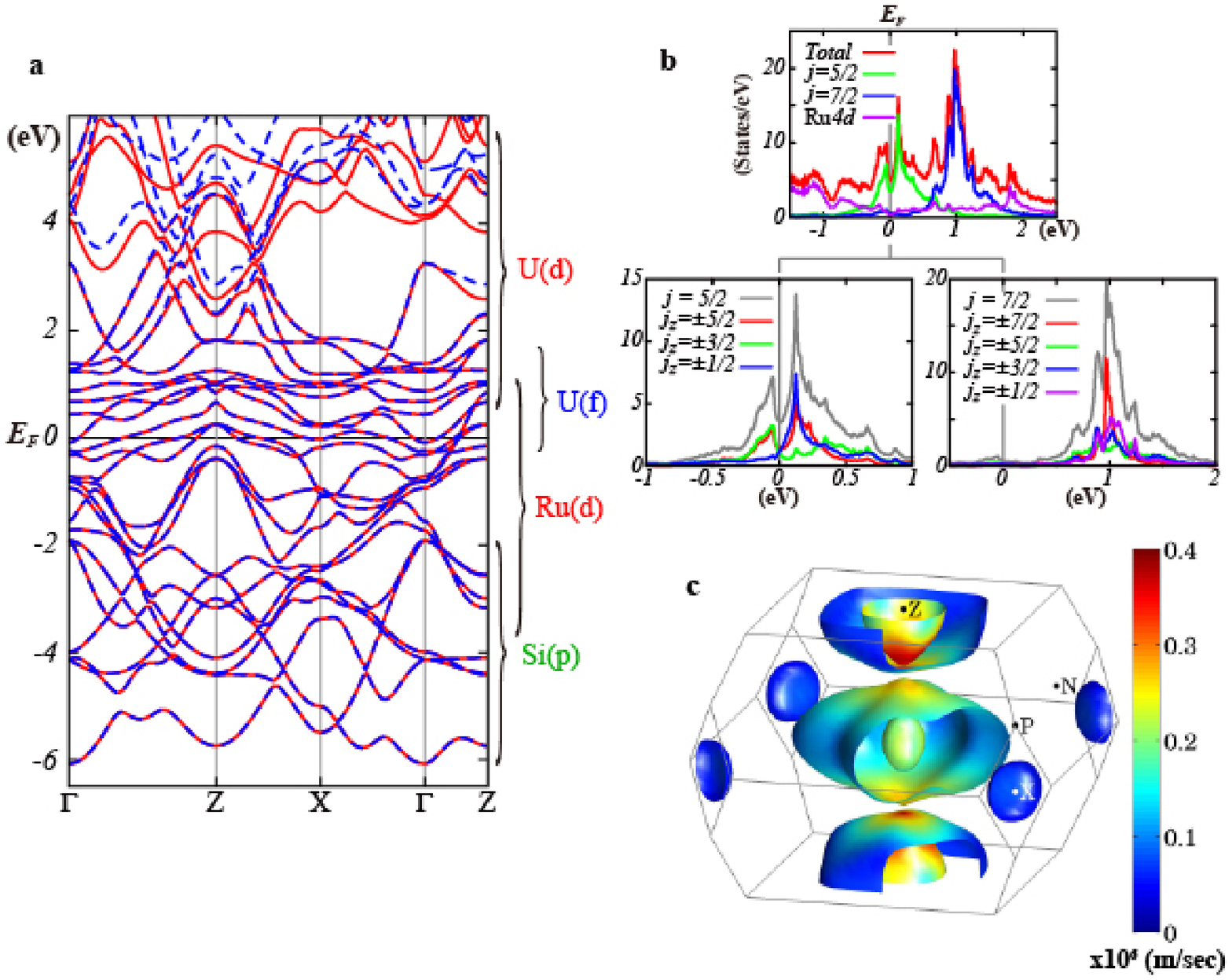}
\caption{Band structure ({\bf a}) along high-symmetry line. Red line is the result of DFT calculation by {\sc wien2k}. Blue dashed line is the Wannier fit. The dispersion below $\sim 2$eV is reproduced completely.  The $J$-resolved DOS ({\bf b}). The left (right) down figure depicts the partial DOS in the $j=5/2$ ($7/2$) manifold.  The Fermi level is located in a dip structure in $j=5/2$ bands. $j=7/2$ bands lies $1$eV higher.  The Fermi surface colored by the Fermi velocity ({\bf c}).  The Fermi velocity is large around $k_z$ axis. }
\end{figure}
The band structure is composed of U $5f$, U $6d$, Ru $4d$ and Si $3p$ orbitals.  U $5f$ band is located where the bottom of U $6d$ band overlap with the top of Ru $4d$ band.  On the basis of the LS basis set, spanned by these orbitals and spin ($\uparrow$ and $\downarrow$), we carry out {\sc wannier90} code \cite{rf:wannier90} via {\sc wien2wannier} interface \cite{rf:w2w}.  With one-shot calculation of the {\sc wannier90}, we obtain a real-space representation of Kohn-Sham equation, i.e., the tight-binding Hamiltonian in $14+10+10\times 2+6\times 2=56$ orbital bases.  We do not make maximally-localized Wannier functions (MLWFs) to preserve the on-site symmetry of the $p$, $d$, and $f$ orbitals, since construction of MLWFs leads to a mixing between up and down spin components at each MLWF.  The obtained Wannier fit (blue dashed line) is well consistent with the original band structure.  The tight-binding Hamiltonian is written as 
\begin{equation}
\begin{split}
H_0=\sum_k\Biggl\{
&\sum_{\ell m}^f E^f_{\boldsymbol{k}\ell m}f^\dagger_{\boldsymbol{k}\ell}f_{\boldsymbol{k}m}
+\sum_{\ell m}^{\rm cond.}\varepsilon_{\boldsymbol{k}\ell m}c^\dagger_{\boldsymbol{k}\ell}c_{km} \\
&~~~~~~+\Biggl( \sum_\ell^{\rm cond.}\sum_m^f V_{\boldsymbol{k}\ell m}c^\dagger_{\boldsymbol{k}\ell}f_{\boldsymbol{k}m}+h.c.\Biggr)
\Biggr\},
\end{split}
\hspace{-3mm}
\end{equation}
where $f_{\boldsymbol{k}\ell}$ ($c_{\boldsymbol{k}\ell}$) is an annihilation operator for an $f$- (conduction-) electron with momentum $\boldsymbol{k}$ and orbital $\ell$.  Superscripts $f$ and $\rm cond.$ in the sums denote all 14 $f$-orbitals and 42 conduction bands, respectively.  The average Wannier spread of each orbital is 0.9, 2.9, 1.6, 3.0 in the atomic unit for U $5f$, U $6d$, Ru $4d$, Si $3p$.  As expected, U $5f$-orbital wavefunctions are well confined at an Uranium site.  This implies that the on-site Coulomb interactions between $f$-electrons is the largest and the most important.  The $f$-orbitals in the LS basis are transformed into the J basis with a unitary matrix, $c_{\ell j}$;
\[
\begin{array}{lll}
c_{4,1}=-\sqrt{1/14}, & c_{5,1}=-\sqrt{1/14}i, & c_{13,1}=-\sqrt{3/7}, \\
c_{14,1}=-\sqrt{3/7}i, & c_{2,2}=\sqrt{1/7}, & c_{3,2}=\sqrt{1/7}i, \\
c_{11,2}=\sqrt{5/14}, & c_{12,2}=\sqrt{5/14}i, & c_{1,3}=-\sqrt{3/7}, \\
c_{9,3}=-\sqrt{2/7}, & c_{10,3}=-\sqrt{2/7}i, & c_{2,4}=-\sqrt{2/7}, \\
c_{3,4}=\sqrt{2/7}i, & c_{8,4}=\sqrt{3/7}, & c_{4,5}=-\sqrt{5/14}, \\
c_{5,5}=\sqrt{1/7}i, & c_{9,5}=\sqrt{1/7}, & c_{10,5}=-\sqrt{1/7}i, \\
c_{6,6}=-\sqrt{3/7}, & c_{7,6}=\sqrt{3/7}i, & c_{11,6}=\sqrt{1/14}, \\
c_{12,6}=-\sqrt{1/14}i, & c_{6,7}=-\sqrt{1/2}, & c_{7,7}=-\sqrt{1/2}i, \\
c_{4,8}=\sqrt{3/7}, & c_{5,8}=\sqrt{3/7}i, & c_{13,8}=-\sqrt{1/14}, \\
c_{14,8}=-\sqrt{1/14}i, & c_{2,9}=-\sqrt{5/14}, & c_{3,9}=-\sqrt{5/14}i, \\
c_{11,9}=\sqrt{1/7}, & c_{12,9}=\sqrt{1/7}i, & c_{1,10}=\sqrt{4/7}, \\
c_{9,10}=-\sqrt{3/14}, & c_{10,10}=-\sqrt{3/14}i, & c_{2,11}=\sqrt{3/14}, \\
c_{3,11}=-\sqrt{3/14}i, & c_{8,11}=\sqrt{4/7}, & c_{4,12}=\sqrt{1/7}, \\
c_{5,12}=-\sqrt{1/7}i, & c_{9,12}=\sqrt{5/14}, & c_{10,12}=-\sqrt{5/14}i, \\
c_{6,13}=\sqrt{1/14}, & c_{7,13}=\sqrt{1/14}i, & c_{11,13}=\sqrt{3/7}, \\
c_{12,13}=-\sqrt{3/7}i, & c_{13,14}=\sqrt{1/2}, & c_{14,14}=-\sqrt{1/2}i,
\end{array}
\]
where subscripts $\ell$ and $j$ denote orbitals in the LS and J bases, respectively.  The LS basis set is given by the direct product of the orbital space, $\{z^3,xz^2,yz^2,z(x^2-y^2),xyz,x(x^2-3y^2),y(3x^2-y^2)\}$, and the spin space $(\uparrow,\downarrow)$.  The J basis set is the direct sum of the total angular momentum $j=5/2$ space, {\it (+5/2,+3/2,+1/2,-1/2,-3/2,-5/2)}, and the $j=7/2$ space, {\it (+7/2,+5/2,+3/2,+1/2,-1/2,-3/2,-5/2,-7/2)}.  

Figures S1{\bf b} shows the $J$-resolved density of states (DOS).   The DOS near the Fermi level, $E_F=0$eV, is dominated by the $j=5/2$ multiplet.  The weight of $j=7/2$ orbitals exists around $1$eV higher due to the local crystalline electric field and the strong spin-orbit interaction, $\lambda L\cdot S$ ($\lambda=0.24$eV).  It should be noted that the Fermi level lies at a deep dip in the DOS.  As shown latter, this feature is important for the temperature dependence of the uniform magnetic susceptibility.  The occupation numbers for these $j=5/2$ and $7/2$ multiplets are 2.07 and 0.64, respectively.  The orbital occupancy in the $j=5/2$ multiplet is 0.72, 0.86, and 0.49 for $J_z=\pm 5/2$, $\pm 3/2$, and $\pm 1/2$, respectively.  

Figure S1{\bf c} is the Fermi surface (FS) colored by the Fermi velocity.  Blue color means heavy band mass. The FS around $k_z$ axis possesses light band mass due to a large mixing with Ru $4d$ bands.  

Finally, we construct a realistic Anderson lattice model by adding the on-site Coulomb interactions between $f$-electrons to this {\it ab initio} tight-binding Hamiltonian, $H_0$.
The interaction parameters $(U,U',J,J')$ are introduced in a conventional form in the LS basis.
\begin{subequations}
\begin{alignat}{1}
H'&=\frac{U}{2}\sum_{i\ell}\sum_{\sigma}f^\dag_{i\ell\sigma}f^\dag_{i\ell\bar\sigma}f_{i\ell\bar\sigma}f_{i\ell\sigma} \\
&+\frac{U'}{2}\sum_{i\ell\ne m}\sum_{\sigma\sigma'}f^\dag_{i\ell\sigma}f^\dag_{im\sigma'}f_{im\sigma'}f_{i\ell\sigma} \\
&+\frac{J}{2}\sum_{i\ell\ne m}\sum_{\sigma\sigma'}f^\dag_{i\ell\sigma}f^\dag_{im\sigma'}f_{i\ell\sigma'}f_{im\sigma} \\
&+\frac{J'}{2}\sum_{i\ell\ne m}\sum_{\sigma}f^\dag_{i\ell\sigma}f^\dag_{i\ell\bar\sigma}f_{im\bar\sigma}f_{im\sigma},
\end{alignat}
\end{subequations}
where $\sigma=\pm$ and $\bar\sigma=-\sigma$.  We measure these interaction parameters in units of $1/\rho_f$, where $\rho_f=4.20$ (States/eV) is the total f-electron DOS at $E_F$.  Hereafter, to account for the mass renormalization effect, the energy and temperature scale is reduced by a factor of 10.

\section{Group-theoretical argument}
We here define one-particle operators for multipole moments from the Group-theoretical argument.  Since the DOS near the Fermi level is dominated by the $j=5/2$ multiplet, we neglect the higher-level $j=7/2$ multiplet, and consider only the $j=5/2$ manifold.  In the six states ($j_z=\pm5/2$, $\pm3/2$, $\pm1/2$), the irreducible tensors are completely available up to rank 5; dipole (rank 1), quadrupole (rank 2), octupole (rank 3), hexadecapole (rank 4), and dotriacontapole (rank 5) moments.  

\begin{table*}
\caption{Definition of multipole moments under the tetragonal $D_{4h}$ symmetry in the $j=5/2$ subspace, which are completely described with 36 bases up to rank 5; dipole ($\boldsymbol{J}$, rank 1), quadrupole ($\boldsymbol{O}$, rank 2), octupole ($\boldsymbol{T}$, rank 3), hexadecapole ($\boldsymbol{H}$, rank 4), and dotriacontapole ($\boldsymbol{D}$, rank 5), including monopole (rank 0). These bases are normalized by Eq.(\ref{eq:normalization}). Superscripts $\pm$ of irreducible representations denote the parity under the time-reversal transformation. $D_{x(y)}\equiv (D_{x(y)a1}+D_{x(y)a2}+D_{x(y)b})/\sqrt{3}$.}
\hspace{-14mm}
\scalebox{0.9}{
\begin{minipage}{\linewidth}
\[
\begin{array}{cll}
\hline {\rm Symmetry} & {\rm Notation} & {\rm Basis} \\
\hline A_1^+ & O_{20} & J_0^{(2)} \\
& H_0 & \bigl(\sqrt{7} J_0^{(4)}+\sqrt{5}\tilde J_{[4]+}^{(4)}\bigr)/\sqrt{12} \\
& H_4 & \bigl(\sqrt{5} J_0^{(4)}-\sqrt{7}\tilde J_{[4]+}^{(4)}\bigr)/\sqrt{12} \\
\hline A_2^+ & H_{za} & \tilde J_{[4]-}^{(4)} \\
\hline B_1^+ & O_{22} & \tilde J_{[2]+}^{(2)} \\
& H_2 & -\tilde J_{[2]+}^{(4)} \\
\hline B_2^+ & O^{xy} & \tilde J_{[2]-}^{(2)} \\
& H_{zb} & \tilde J_{[2]-}^{(4)} \\
\hline E^+ & O_{yz},O_{zx} & \tilde J_{[1]+}^{(2)},\tilde J_{[1]-}^{(2)} \\
& H_{xa},H_{ya} & \bigl( \tilde J_{[3]+}^{(4)}+\sqrt{7}\tilde J_{[1]+}^{(4)}\bigr)/\sqrt{8},-\bigl( \tilde J_{[3]-}^{(4)}-\sqrt{7}\tilde J_{[1]-}^{(4)}\bigr)/\sqrt{8} \\
& H_{xb},H_{yb} & \bigl(\sqrt{7} \tilde J_{[3]+}^{(4)}-\tilde J_{[1]+}^{(4)}\bigr)/\sqrt{8},-\bigl(\sqrt{7} \tilde J_{[3]-}^{(4)}+\tilde J_{[1]-}^{(4)}\bigr)/\sqrt{8} \\
\hline A_1^- & D_4 & \tilde J_{[4]-}^{(5)} \\
\hline A_2^- & J_z & J_0^{(1)} \\
& T_{za} & J_0^{(3)} \\
& D_{za1} & J_0^{(5)} \\
& D_{za2} & \tilde J_{[4]+}^{(5)} \\
\hline B_1^- & T^{xyz} & \tilde J_{[2]-}^{(3)} \\
& D_2 & -\tilde J_{[2]-}^{(5)} \\
\hline B_2^- & T^{zb} & \tilde J_{[2]+}^{(3)} \\
& D_{zb} & \tilde J_{[2]+}^{(5)} \\
\hline E^- & J_x,J_y & \tilde J_{[1]-}^{(1)}, \tilde J_{[1]+}^{(1)} \\
& T_{xa},T_{ya} & \bigl(\sqrt{5}\tilde J_{[3]-}^{(3)}-\sqrt{3}\tilde J_{[1]-}^{(3)}\bigr)/\sqrt{8},
-\bigl(\sqrt{5}\tilde J_{[3]+}^{(3)}+\sqrt{3}\tilde J_{[1]+}^{(3)}\bigr)/\sqrt{8} \\
& T_{xb},T_{yb} & -\bigl(\sqrt{3}\tilde J_{[3]-}^{(3)}+\sqrt{5}\tilde J_{[1]-}^{(3)}\bigr)/\sqrt{8},
-\bigl(\sqrt{3}\tilde J_{[3]+}^{(3)}-\sqrt{5}\tilde J_{[1]+}^{(3)}\bigr)/\sqrt{8} \\
& D_{xa1},D_{ya1} & \bigl( 3\sqrt{14}\tilde J_{[5]-}^{(5)}-\sqrt{70}\tilde J_{[3]-}^{(5)}+2\sqrt{15}\tilde J_{[1]-}^{(5)}\bigr)/16, \bigl( 3\sqrt{14}\tilde J_{[5]+}^{(5)}+\sqrt{70}\tilde J_{[3]+}^{(5)}+2\sqrt{15}\tilde J_{[1]+}^{(5)}\bigr)/16 \\
& D_{xa2},D_{ya2} & \bigl( \sqrt{10}\tilde J_{[5]-}^{(5)}+9\sqrt{2}\tilde J_{[3]-}^{(5)}+2\sqrt{21}\tilde J_{[1]-}^{(5)}\bigr)/16, \bigl( \sqrt{10}\tilde J_{[5]+}^{(5)}-9\sqrt{2}\tilde J_{[3]+}^{(5)}+2\sqrt{21}\tilde J_{[1]+}^{(5)}\bigr)/16 \\
& D_{xb},D_{yb} & \bigl( \sqrt{30}\tilde J_{[5]-}^{(5)}+\sqrt{6}\tilde J_{[3]-}^{(5)}-2\sqrt{7}\tilde J_{[1]-}^{(5)}\bigr)/8, \bigl( \sqrt{30}\tilde J_{[5]+}^{(5)}-\sqrt{6}\tilde J_{[3]+}^{(5)}-2\sqrt{7}\tilde J_{[1]+}^{(5)}\bigr)/8 \\
\hline
\end{array}
\]
\end{minipage}
}
\end{table*}

The irreducible tensor for rank $k$ has $2k+1$ components $J_q^{(k)}$, which meet the following relations,
\begin{align}
\Bigl[J_z,J_q^{(k)}\Bigr] &=qJ_q^{(k)}, \\
\Bigl[J_\pm,J_q^{(k)}\Bigr] &=\sqrt{(k\mp q)(k\pm q+1)}J_{q\pm1}^{(k)}, 
\end{align}
where $J_\pm=J_x\pm i J_y$ are raising and lowering operators, and $J_\mu$ is $\mu$ component of the total angular momentum operator.  In Table S1, we show multipole moments and Hermite bases belonging to each irreducible representation of the tetragonal symmetry ($D_{4h}$) \cite{rf:Shiina}.  Here $\tilde{J}_{[q]\pm}^{(k)}$ is defined by
\begin{subequations}
\begin{align}
\tilde{J}_{[2p-1]+}^{(k)}&=\frac{i}{\sqrt{2}}\Bigl( J_{2p-1}^{(k)}+J_{-(2p-1)}^{(k)} \Bigr), \\
\tilde{J}_{[2p-1]-}^{(k)}&=\frac{1}{\sqrt{2}}\Bigl( -J_{2p-1}^{(k)}+J_{-(2p-1)}^{(k)} \Bigr), \\
\tilde{J}_{[2p]+}^{(k)}&=\frac{1}{\sqrt{2}}\Bigl( J_{2p}^{(k)}+J_{-2p}^{(k)} \Bigr), \\
\tilde{J}_{[2p]-}^{(k)}&=\frac{i}{\sqrt{2}}\Bigl( -J_{2p}^{(k)}+J_{-2p}^{(k)} \Bigr),
\end{align}
\end{subequations}
for positive integer, $p$.  Applying the Wigner-Eckart theorem, we obtain the matrix elements of $J_q^{(k)}$ as
\begin{equation}
\bigl<jj_z|J_q^{(k)}|jj'_z\bigr>=\bigl<j||J^{(k)}||j\bigr>\frac{\bigl<jj_z|jj'_zkq\bigr>}{\sqrt{2j+1}},
\end{equation}
where $\bigl<j||J^{(k)}||j\bigr>$ is the reduced matrix element, and $\bigl<jj_z|jj'_zkq\bigr>$ is the Clebsch-Gordan coefficients. These are easily evaluated in terms of {\sc Mathematica}.  Thus we obtain numerically all representation matrices in Table S1.  For example, matrix elements of higher-rank multipoles, $H_{xb}$, $D_4$, and $D_{x(y)}=(D_{x(y)a1}+D_{x(y)a2}+D_{x(y)b})/\sqrt{3}$, are explicitly given by
\begin{equation}
H_{xb} = \left(
\begin{array}{rrrrrr}
& 0.09i && \!\!\!-0.47i && 0.00i \cr
\!\!\!-0.09i && \!\!\!-0.15i && \phantom{\!\!\!-}0.00i & \cr
& 0.15i && 0.00i && 0.47i \cr
0.47i && 0.00i && 0.15i & \cr
& \phantom{\!\!\!-}0.00i && \!\!\!-0.15i && \!\!\!-0.09i \cr
0.00i && \!\!\!-0.47i && 0.09i
\end{array}
\!\right)\!,
\end{equation}

\begin{equation}
D_4 = \left(
\begin{array}{rrrrrr}
&&&& \!\!\!-0.50i & \cr
&&& 0.00i && \phantom{\!\!\!-}0.50i \cr
&& 0.00i && 0.00i \cr
& 0.00i && \phantom{\!\!\!-}0.00i \cr
\phantom{\!\!\!-}0.50i && \phantom{\!\!\!-}0.00i \cr
& \!\!\!-0.50i
\end{array}
\!\right)\!,
\end{equation}

\begin{equation}
D_x= \left(
\begin{array}{rrrrrr}
& \!\!\!0.02 && 0.11 && \!\!\!0.65 \cr
0.02 && -0.08 && \!\!\!-0.08 & \cr
& \!\!\!-0.08 && \!\!\!0.11 && 0.11 \cr
\!\!\!0.11 && 0.11 && -0.08 & \cr
& -0.18 && \!\!\!-0.08 && \!\!\!0.02 \cr
0.65 && \!\!\!0.11 && 0.02
\end{array}
\!\right)\!, 
\end{equation}

\begin{equation}
D_y= \left(
\begin{array}{rrrrrr}
& \!\!\!-0.02i && 0.11i && \!\!\!-0.65i \cr
0.02i && 0.08i && \!\!\!-0.08i & \cr
& \!\!\!-0.08i && \!\!\!-0.11i && 0.11i \cr
\!\!\!-0.11i && 0.11i && 0.08i & \cr
& 0.18i && \!\!\!-0.08i && \!\!\!-0.02i \cr
0.65i && \!\!\!-0.11i && 0.02i
\end{array}
\!\right)\!, 
\end{equation}
where those norms are normalized by Eq.(\ref{eq:normalization}).

\section{RPA and Magnetic correlations}
\subsection{Formalism}
First, we calculate one-particle Green functions in the LS basis,
\begin{subequations}
\begin{align}
G_{\ell m}(\boldsymbol{k},i\omega_n)& =-\bigl<\!\bigl<f_{\boldsymbol{k}\ell} f_{\boldsymbol{k}m}^\dagger \bigr>\!\bigr> \\
&= -\!\int_0^\beta d\tau e^{i\omega_n\tau} \bigl< T_\tau [f_{\boldsymbol{k}\ell}(\tau)f_{\boldsymbol{k}m}^\dagger(0)] \bigr>
\end{align}
\end{subequations}
where $\ell$ and $m$ denote both $f$-orbital and spin quantum number.  The non-interacting static full susceptibility is given by
\begin{equation}
\chi^0_{\ell m,\ell' m'}(\boldsymbol{q},0)=-T\sum_{\boldsymbol{k},n}G_{\ell\ell'}(\boldsymbol{k},i\omega_n)G_{m'm}(\boldsymbol{k}+\boldsymbol{q},i\omega_n).
\end{equation}
This is written in the $14^2 \times 14^2$ matrix form, $\hat{\chi}^0(q)$, with the $\{\ell m\}$ row and the $\{\ell'm'\}$ column.  In this case, the RPA susceptibility is given by
\begin{subequations}
\begin{align}
\hat\chi^{\rm RPA}(q)&=\hat\chi^0(q)+\hat\chi^0(q)\hat\Gamma^0\hat\chi^{\rm RPA}(q) \\
&=\left[ 1-\hat\chi^0(q)\hat\Gamma^0 \right]^{-1}\hat\chi^0(q),
\end{align}
\end{subequations}
where $\Gamma^0$ is the bare interactions between $f$-electrons, 
\[
\begin{array}{lllc}
\Gamma^0_{\ell\ell,mm}&=\Gamma^0_{\ell+7\ell+7,m+7 m+7}&=&J-U', \\
\Gamma^0_{\ell m,\ell m}&=\Gamma^0_{\ell+7 m+7,\ell+7 m+7}&=&U'-J, \\
\Gamma^0_{\ell\ell,m+7 m+7}&=\Gamma^0_{\ell+7 \ell+7,m m}&=&-U', \\
\Gamma^0_{\ell m,m+7 \ell+7}&=\Gamma^0_{\ell+7 m+7,m \ell}&=&-J', \\
\Gamma^0_{\ell m,\ell+7 m+7}&=\Gamma^0_{\ell+7 m+7,\ell m}&=&-J, \\
\Gamma^0_{\ell \ell+7,m m+7}&=\Gamma^0_{\ell+7 \ell, m+7 m}&=&J, \\
\Gamma^0_{\ell m+7,m \ell+7}&=\Gamma^0_{\ell+7 m,m+7 \ell}&=&J', \\
\Gamma^0_{\ell m+7,\ell m+7}&=\Gamma^0_{\ell+7 m,\ell+7 m}&=&U', \\
\Gamma^0_{\ell,\ell,\ell+7,\ell+7}&=\Gamma^0_{\ell+7,\ell+7,\ell,\ell}&=&-U, \\
\Gamma^0_{\ell,\ell+7,\ell,\ell+7}&=\Gamma^0_{\ell+7,\ell,\ell+7,\ell}&=&U,
\end{array}
\]
with orbital indexes $\ell \ne m \in 1 \sim 7$.
Correlation between multipole $A$ and multipole $B$ is evaluated by the product of the above susceptibility and the representation matrices, 
\begin{equation}
\bigl<\!\bigl< A, B \bigr>\!\bigr>=\sum_{\ell\ell'mm'} A_{m \ell}\chi^{\rm RPA}_{\ell m,\ell' m'}(\boldsymbol{q})B_{\ell'm'}.
\end{equation}
In the LS basis set, the spin moment $S_\mu$ with $\mu=x,y,z$ is the product of the unit matrix in the orbital space and the Pauli matrices in the spin space, $\hat1\otimes\hat\sigma^\mu/2$.  The orbital moment is the product of the unit matrix in the spin space and $L_\mu$;
\begin{equation}
L_x=\!\left(
\begin{array}{ccccccc}
0&& \sqrt{6}i &&&&\\
&0&&& \frac{\sqrt{10}}{2}i &&\\
\!\!\!-\sqrt{6}i &&0& \!\!\!-\frac{\sqrt{10}}{2}i &&&\\
&& \frac{\sqrt{10}}{2}i &0&&& \frac{\sqrt{6}}{2}i \\
& \!\!\!-\frac{\sqrt{10}}{2}i &&&0& \!\!\!-\frac{\sqrt{6}}{2}i &\\
&&&& \frac{\sqrt{6}}{2}i &0&\\
&&& \!\!\!-\frac{\sqrt{6}}{2}i &&&0
\end{array}
\!\!\right)\!,
\end{equation}
\begin{equation}
L_y=\!\left(
\begin{array}{ccccccc}
0& \!\!\!-\sqrt{6}i & 0 \\
\sqrt{6}i &0&& \!\!\!-\frac{\sqrt{10}}{2}i \\
0&&0&& \!\!\!-\frac{\sqrt{10}}{2}i \\
& \frac{\sqrt{10}}{2}i &&0&& \!\!\!-\frac{\sqrt{6}}{2}i \\
&& \frac{\sqrt{10}}{2}i &&0&& \!\!\!-\frac{\sqrt{6}}{2}i \\
&&& \frac{\sqrt{6}}{2}i &&0 \\
&&&& \frac{\sqrt{6}}{2}i && 0
\end{array}
\!\!\right)\!,
\end{equation}
\begin{equation}
L_z=\left(
\begin{matrix}
&0 \\
0&& -i \\
& i &&0 \\
&&0&& -2i \\
&&& 2i && 0 \\
&&&&0&& -3i\\
&&&&& 3i
\end{matrix}
\right).
\end{equation}
The total magnetic moment $M_\mu$ is defined by $L_\mu+2S_\mu$.

Next, let us consider only the $j=5/2$ subspace, neglecting the $j=7/2$ subspace.  The magnetic moment is approximated as $M_\mu=gJ_\mu$, where $g=6/7$ is the Lande $g$-factor.  The magnetic dipole moment, $J_\mu$, in the $j=5/2$ subspace is given by

\begin{equation}
J_x=\left(
\begin{array}{cccccc}
0& \frac{\sqrt{5}}{2} \cr
\frac{\sqrt{5}}{2} &0& \sqrt{2} \cr
& \sqrt{2} &0& \frac{3}{2} \cr
&& \frac{3}{2} &0& \sqrt{2} \cr
&&& \sqrt{2} &0& \frac{\sqrt{5}}{2} \cr
&&&& \frac{\sqrt{5}}{2} &0 
\end{array}
\right),
\end{equation}

\begin{equation}
J_y=\left(
\begin{array}{cccccc}
0& -\frac{\sqrt{5}}{2}i  \cr
\frac{\sqrt{5}}{2}i &0& -\sqrt{2}i  \cr
& \sqrt{2}i &0& -\frac{3}{2}i  \cr
&& \frac{3}{2}i &0& -\sqrt{2}i  \cr
&&& \sqrt{2}i &0& -\frac{\sqrt{5}}{2}i \cr
&&&& \frac{\sqrt{5}}{2}i &0 
\end{array}
\right),
\end{equation}

\begin{equation}
J_z=\left(
\begin{array}{cccccc}
\frac{5}{2}  \cr
& \frac{3}{2}  \cr
&& \frac{1}{2}  \cr
&&& -\frac{1}{2}  \cr
&&&& -\frac{3}{2}  \cr
&&&&& -\frac{5}{2} 
\end{array}
\right).
\end{equation}
In actual calculations, since we have already obtained the RPA susceptibility in the LS bases, we need to transfer these J-base representation matrices into the LS-basis ones, $\sum_{jj'}c_{\ell j}J_\mu^{jj'}c_{mj'}^*$ by the unitary matrix $c_{\ell j}$.

\subsection{Magnetic correlations}
We carry out the RPA analysis for several parameters in $32\times 32\times 8$ $k$-meshes.  For simplicity, we show in the present study only some results for $U=U'\simeq 2.3$ and $J=J'=0$.  The obtained characteristic features are barely changed for $J=J'>0$.  Mainly, the RPA treatment remarkably enhances the magnetic character in the non-interacting system.  Figure S2 is the magnetic correlations along high-symmetry line.  Figure S2{\bf a} demonstrates a remarkable magnetic anisotropy, namely, the $L_z$ correlation $\bigl<\!\bigl<L_z,L_z\bigr>\!\bigr>$ larger than the $L_{x(y)}$ correlations $\bigl<\!\bigl<L_{x(y)},L_{x(y)}\bigr>\!\bigr>$.  The $L_z$ correlation indicates a peak structure at $Z$ (1 0 0) point and a hump structure at (0.6 0 0), while the $L_{x(y)}$ correlation is featureless.  In these correlations, the correlation between orbital moments (red line) are much larger than spin-spin correlations (green line).   The correlations between spin and orbital moments (blue line) are negative, and compensate the large orbital-orbital correlations.  This means that the large orbital moment is compensated by the anti-parallel spin moment.  Generally, this is the case in the Uranium compounds, since the electron occupation of $f$ orbitals is less than half.  In Fig.~S2{\bf b}, we compare the total magnetic correlations with the $gJ$ correlations confined in the $j=5/2$ subspace.  As expected from the fact that the DOS near $E_F$ is dominated by the $j=5/2$ components, the total magnetic correlations are overall explained by the $j=5/2$ correlations.  Thus we expect that we can neglect the effect of the $j=7/2$ subspace also in the higher-rank multipoles.

\begin{figure}[h]
\includegraphics[width=\textwidth]{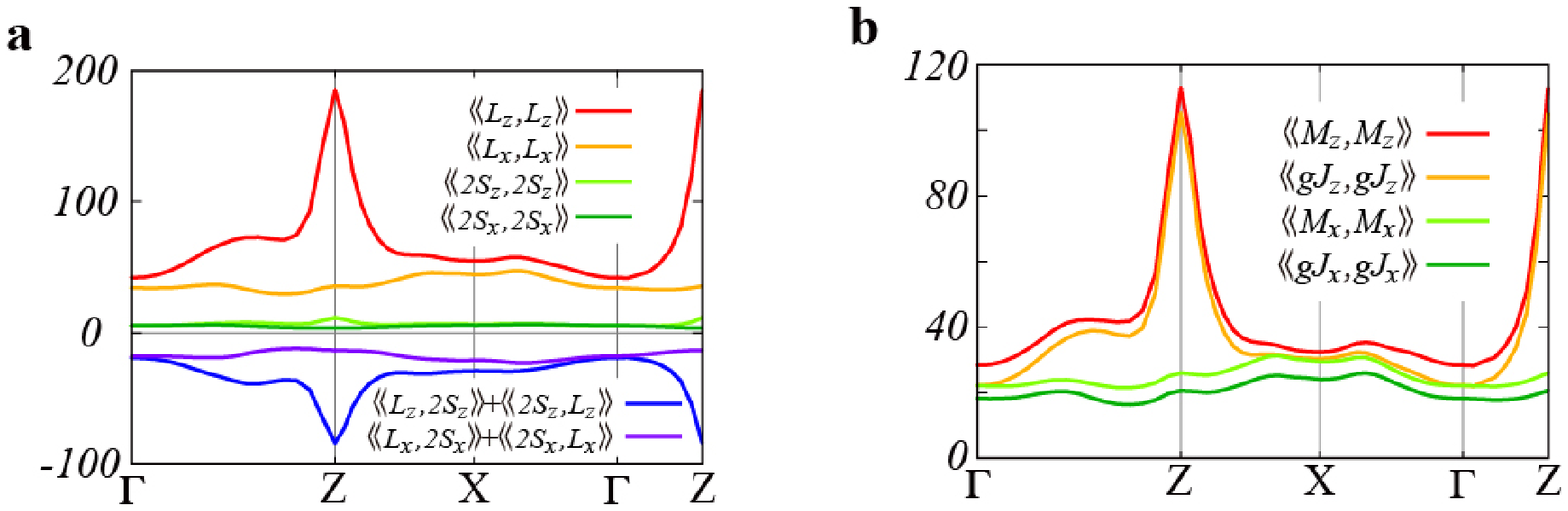}
\caption{LS-resolved magnetic correlations ({\bf a}) along high-symmetry line, where $M=L+2S$. The orbital contribution is much larger than spin contribution (green line). Correlation between orbital and spin (blue line) is negative due to the less than half.  $j=5/2$ subspace contribution ({\bf b}) to the total magnetic correlations along high-symmetry line. Differences between $\bigl<\!\bigl<M_\mu,M_\mu\bigr>\!\bigr>$ correlation and $g^2\bigl<\!\bigl<J_\mu,J_\mu\bigr>\!\bigr>$ correlation come from the contribution of $j=7/2$ manifold, which provides only small constant shift.}
\end{figure}

\begin{figure}[h]
\includegraphics[height=55mm]{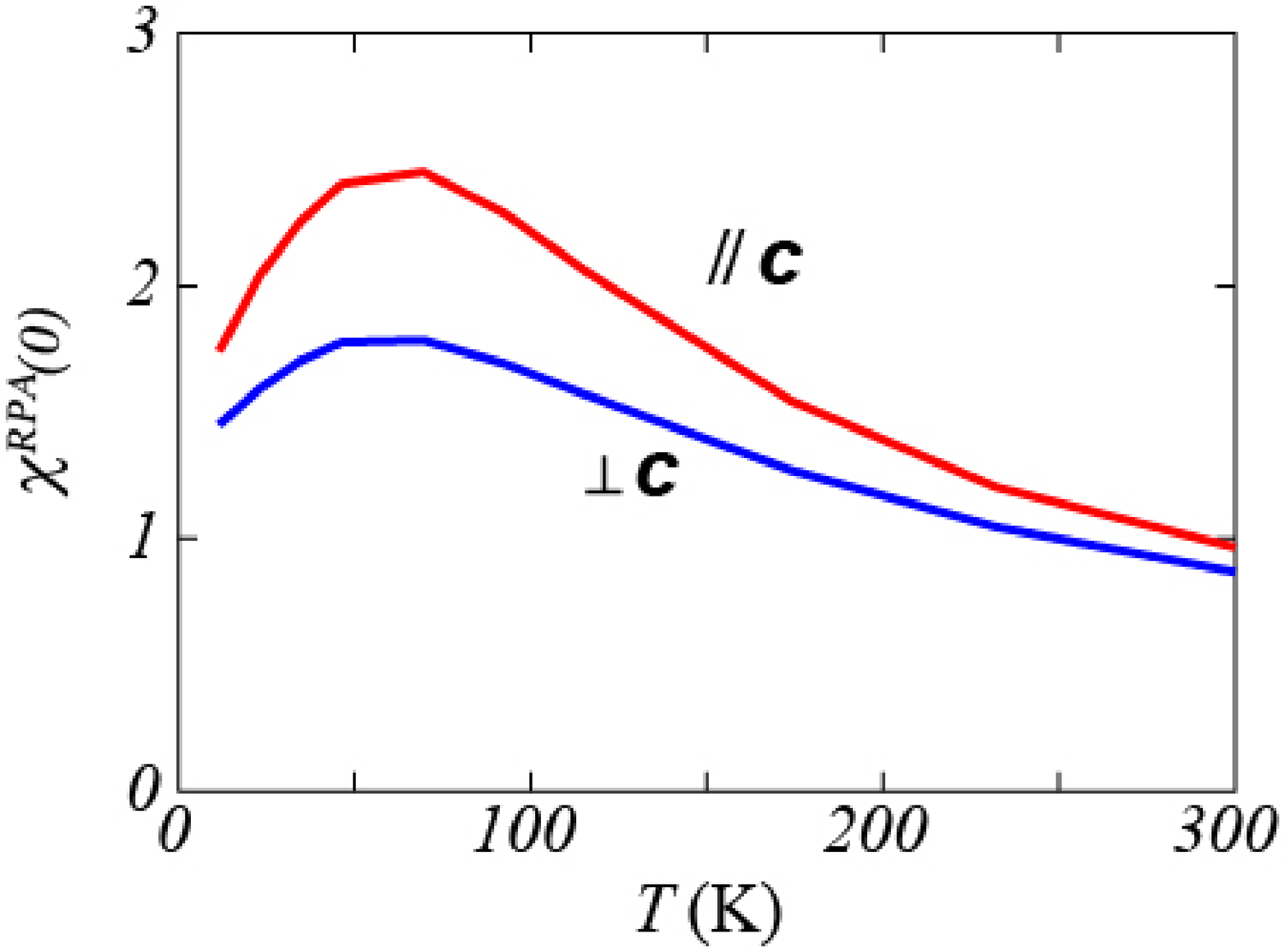}
\caption{Temperature dependence of the uniform susceptibility within the RPA.  The $J_z$ correlation parallel to the $c$ axis (red), $\chi_c^{\rm RPA}(0)$, is larger than the in-plane $J_{x(y)}$ correlation (blue), $\chi_{ab}^{\rm RPA}(0)$.  Both correlations show a broad maximum at around 50K. }
\end{figure}

Figure S3 depicts temperature dependence of the uniform magnetic correlations within the RPA.  
The $J_z$ correlation $\chi_c^{\rm RPA}(0)$ parallel to the $c$ axis is larger than the in-plane $J_{x(y)}$ correlation $\chi_{ab}^{\rm RPA}(0)$.  Both correlations show a broad maximum at around 50K.  Such temperature dependence is similar to the experimental result for the $c$-axis magnetic susceptibility.  This feature arises from the fact that the DOS possesses a deep dip structure near $E_F$.  In calculations beyond the RPA (Sec.~V), the uniform susceptibility $\chi_c(0)$ parallel to the $c$ axis is enhanced, while the in-plane $\chi_{ab}(0)$ is suppressed. In this case we obtain more remarkable Ising-like anisotropy including the temperature-independent behavior in $\chi_{ab}(0)$ (Fig.~3{\bf c}).

\section{Multipole correlations}
We obtain higher-rank multipole correlations (Fig.~2) in the same way as the $J_\mu$ correlations.  In order to compare different-rank multipole correlations, we here normalize those norms as
\begin{equation}
\sum_{\ell m}|Q_{\ell m}|^2=1,  \label{eq:normalization}
\end{equation}
where $Q_{\ell m}$ represents matrix elements of multipole moments in Table S1.  Owing to this normalization, magnitude of dipole correlations in Fig.~2{\bf a} becomes $2/(35g^2)$ smaller than in Fig.~S2{\bf b}.  Figure 2 shows the following features.  In quadrupole correlations, $O_{20}$ and $O_{yz(zx)}$ are dominant at $Z$ point.  Octupole correlations have featureless $Q$ dependence.  In hexadecapole correlations, $H_{x(y)\beta}$ is dominant.  In dotriacontapole correlations, $D_4$, $D_{z\alpha2}$, $D_{x(y)\alpha1,2}$ and $D_{x(y)\beta}$ show a peak structure at $Z$ point.  It should be noted that the off-diagonal part (red lines) is large in dotriacontapole correlations.  This means that different bases have a large mixing.  Such mixing is not restricted in the same rank.  Thus we need to diagonalize the full susceptibility $\chi_{\ell m,\ell'm'}(\boldsymbol{Q}_C)$ itself in the $14^2\times 14^2$ matrix form.  The results were shown in Fig.~2{\bf g}.  The obtained $n$th eigenvector, $\Psi^n_{\ell m}$, can be expanded in terms of 36 multipoles, $Q^j_{\ell m}$, shown in Table S1;
\begin{equation}
\Psi^n_{\ell m}=\sum_j \Delta^n_j Q^j_{\ell m}.
\end{equation}
Each component of $\Delta^n_j$ was obtained as the following,
\begin{equation}
\begin{array}{llll}
\Psi^1_{A_2^-}&\!\!\!=&\!\!\phantom{-}0.824J_z+0.281T_{za}-0.298D_{za1}+0.391D_{za2}, \\[1pt]
\Psi^2_{E^-}&\!\!\!=&\!\!\phantom{-}0.109J_y-0.341T_{ya}+0.190T_{yb} \\[1pt]
&&\!\!+0.612D_{ya1}+0.413D_{ya2}+0.540D_{yb}, \\[1pt]
\Psi^{2'}_{E^-}&\!\!\!=&\!\!\phantom{-}0.109J_x-0.341T_{xa}+0.190T_{xb} \\[1pt]
&&\!\!+0.612D_{xa1}+0.413D_{xa2}+0.540D_{xb}, \\[1pt]
\Psi^3_{A_1^-}&\!\!\!=&\!\!\phantom{-1.000}D_4, \\[1pt]
\Psi^4_{E^+}&\!\!\!=&\!\!-0.266O_{yz}-0.372H_{xa}+0.889H_{xb}, \\[1pt]
\Psi^{4'}_{E^+}&\!\!\!=&\!\!-0.266O_{zx}-0.372H_{ya}+0.889H_{yb}, \\[1pt]
\Psi^5_{B_2^+}&\!\!\!=&\!\!-0.291O_{xy}+0.957H_{zb}.
\end{array}
\end{equation}
$E^-$ and $E^+$ representations are twofold degenerate, $\Psi^{2(2')}_{E^-}$ and $\Psi^{4(4')}_{E^-}$.  The dominant order parameter is $A_2^-$ dipole, $E^-$ dotriacontapole, $A_1^-$ dotriacontapole, and $E^+$ hexadecapole, and $B_2^+$ hexadecapole, in order.  The $E^-$ state includes comparably $D_{x(y)a1}$, $D_{x(y)a2}$, and $D_{x(y)b}$.  Thus we have introduced a new bases $D_{x(y)}=(D_{x(y)a1}+D_{x(y)a2}+D_{x(y)b})/\sqrt{3}$ in the present study.

\section{beyond RPA and Psudospin representations}
\subsection{beyond RPA}
Generally, RPA enhances a magnetic channel, but depresses a charge channel.  This trend is improved by including higher-order fluctuations beyond the RPA, i.e., the mode-mode coupling terms such as Maki-Thompson type vertex corrections.  In multi-orbital systems, such mode-mode coupling term mix also many different channels.  This implies that a channel which is not enhanced within the RPA may develop.  Therefore calculations beyond the RPA is very important especially in the multi-orbital systems.  We here examine Maki-Thompson type mode-mode coupling  effect to search a possibility that some high-rank multipole correlation overcomes the dipole $J_z$ correlation, which is dominantly enhanced within the RPA.  

\begin{figure*}[h]
\includegraphics[width=0.8\textwidth]{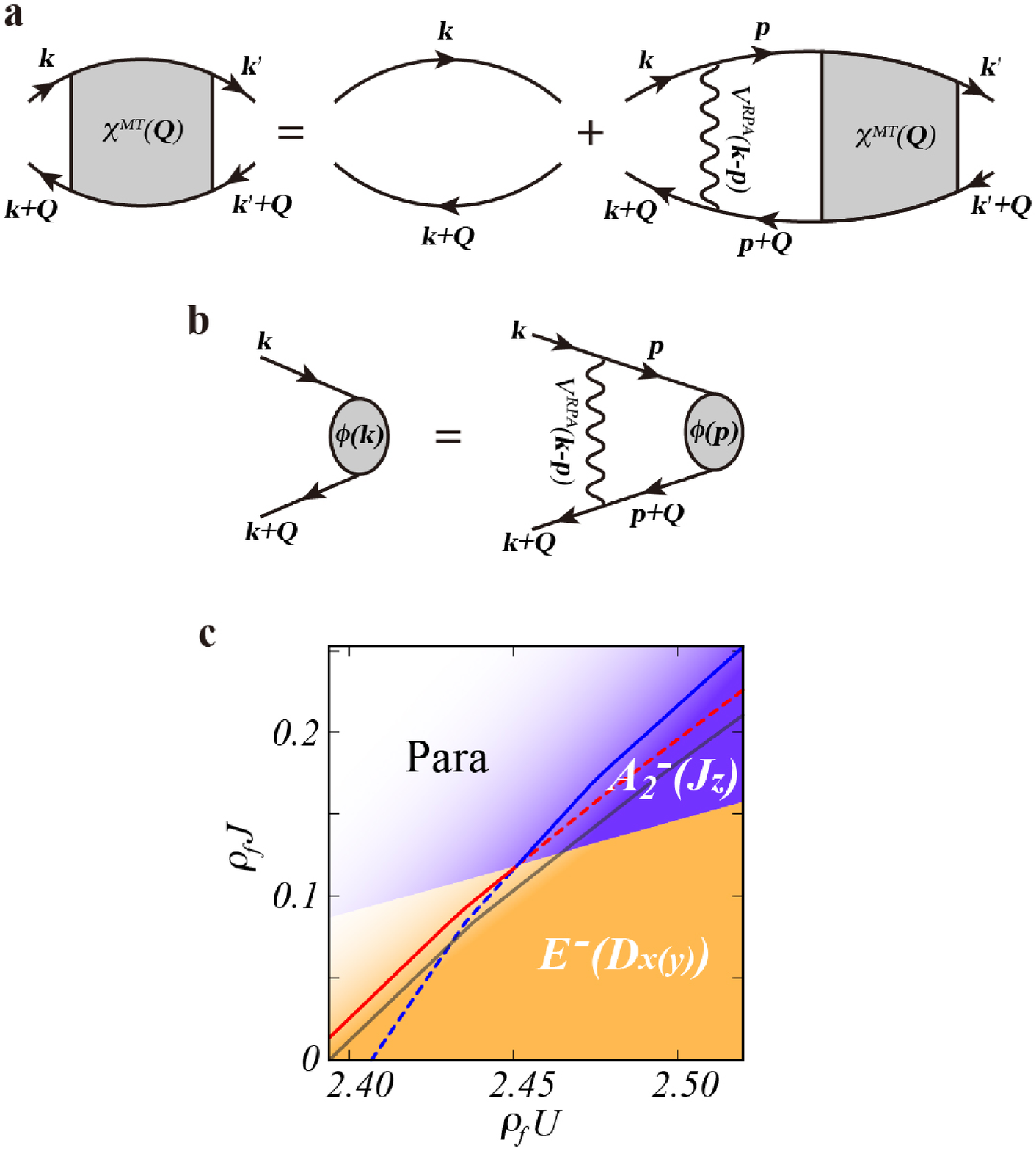}
\caption{Maki-Thompson type diagrams ({\bf a}) and the corresponding Bethe-Salpheter equation ({\bf b}).  $U-J$ phase diagram for $U=U'+2J$ and $J=J'$ at $T=16$K ({\bf c}).  $E^-$ and $A_2^-$ states fulfill $\lambda=1$ along red and blue lines, respectively. Broken lines are not realized.  Darker background color corresponds to larger eigenvalue.  Grey line denotes a change in the parameter set used in Fig.~3{\bf b}.}
\end{figure*}

The present Maki-Thompson type diagrams are shown in Fig.~S4{\bf a}.  The wavy line represents an effective interaction including the RPA results,
\begin{equation}
V_{\ell m,\ell'm'}(\boldsymbol{q}) = \Gamma_{\ell m,\ell'm'}^0+\sum_{\ell_1m_1\ell_2m_2}\Gamma_{\ell m,\ell_1m_1}^0\chi^{\rm RPA}_{\ell_1 m_1,\ell_2 m_2}(\boldsymbol{q},0)\Gamma_{\ell_2 m_2,\ell'm'}^0.
\end{equation}
Here we include only a static part $\nu_n=0$ of $\chi^{\rm RPA}_{\ell_1 m_1,\ell_2 m_2}(\boldsymbol{q},i\nu_n)$, which corresponds to the classical approximation.  In this case, it is convenient to calculate not diagrams in Fig.~S4{\bf a}, but the maximum eigenvalue $\lambda$ of the Bethe-Salpheter equation (Fig.~S4{\bf b}), 
\begin{equation}
\lambda \phi_{\ell m}(\boldsymbol{k}) = V_{\ell \ell_1,m m_1}(\boldsymbol{k}-\boldsymbol{p}) T\sum_n G_{\ell_1\ell'}(\boldsymbol{p},i\omega_n)G_{m'm_1}(\boldsymbol{p}+\boldsymbol{Q}_C,i\omega_n) \phi_{\ell'm'}(\boldsymbol{p}),
\end{equation}
where $\phi_{\ell m}(\boldsymbol{k})=\langle f^\dagger_{\boldsymbol{k}\ell}f_{\boldsymbol{k}+\boldsymbol{Q}_C m}\rangle$ is the order parameter for the staggered pair, which has generally the momentum dependence.  Indeed such $k$-dependent order parameters have been proposed in some scenarios, including the unconventional spin density wave~\cite{rf:Ikeda}, the orbital AFM~\cite{rf:Chandra}, the Helicity order~\cite{rf:Varma}, and the spin nematic~\cite{rf:Fujimoto}.  However, the present calculations indicate that $\phi_{\ell m}(k)$ with large eigenvalue $\lambda$ is almost $k$ independent (not shown).  Thus, it is appropriate that such order parameter is considered to be a kind of multipole, which is almost on-site pairing.  The results for $U=U'\simeq 2.4$ and $J=J'=0$ has been shown in Fig.~3{\bf a}.  Following Eq.(S22), $k$-independent parts of the dominant two eigenstates, $E^-$ and $A_2^-$, are expanded as 
\begin{equation}
\begin{array}{llll}
\phi_{E^-}&\!\!=& 0.150J_{x(y)}-0.216T_{x(y)a}+0.206T_{x(y)b} \\[1pt]
&&\!\!\!\!+0.671D_{x(y)a1}+0.296D_{x(y)a2}+0.592D_{x(y)b}, \\[1pt]
\phi_{A_2^-}&\!\!=& 0.825J_z+0.406T_{za}-0.140D_{za1}+0.367D_{za2}.
\end{array}
\end{equation}
Thus the main ingredient of $\phi_{E^-}$ is $D_{x(y)}$, while that of $\phi_{A_2^-}$ is $J_z$.  This is the same as in the RPA results. 

Fig.~S4{\bf c} is a $U-J$ diagram for $U=U'+2J$ and $J=J'$ at $T=16$K.  Large $J$ stabilizes $A_2^-(J_z)$ state rather than $E^-(D_{x(y)})$ state.  Red and blue lines represent parameter lines fulfilled $\lambda=1$ for $A_2^-(J_z)$ and $E^-(D_{x(y)})$, respectively.  Grey line corresponds to the parameter set used in Fig.~3{\bf b}.

Finally, within the same framework, we study the uniform magnetic susceptibility (Fig.~3{\bf c}).  This is obtained as $\boldsymbol{Q}=0$ in Fig.~S4{\bf a},
\begin{subequations}
\begin{align}
& \Pi_{\ell m,\ell'm'}(k) =\Pi_{\ell m,\ell'm'}^0(k)-\Pi_{\ell m,\ell_1m_1}^0(k)V_{\ell_1 \ell_2,m_1 m_2}(k-k')\Pi_{\ell_2 m_2,\ell'm'}(k), \\
& \Pi^0_{\ell m,\ell'm'}(k) = -T\sum_n G_{\ell\ell'}(k,i\omega_n)G_{mm'}(k,i\omega_n), \\
& \chi_\mu(0) =\sum_{k} J_\mu^{m\ell} \Pi_{\ell m,\ell'm'}(k) J_\mu^{\ell'm'}.
\end{align}
\end{subequations}

\subsection{Pseudospin representations}
When we focus on only $\pm 5/2$ orbitals, three multipole moments $(D_x,D_y,J_z)$ defined in eqs.(S9), (S10) and (S20) are reduced to  
\begin{equation}
D_x=\left(\begin{array}{cc}
0 & 0.65 \cr
0.65 & 0 
\end{array}\right), ~~~
D_y=\left(\begin{array}{cc}
0 & 0.65i \cr
-0.65i & 0 
\end{array}\right), ~~~
J_z=\left(\begin{array}{cc}
0.6 & 0 \cr
0 & -0.6
\end{array}\right),
\end{equation}
where these are normalized by eq.(\ref{eq:normalization}).
These are proportional to the Pauli matrices, ($\sigma_x$, $\sigma_y$, $\sigma_z$).  Thus we can describe these moments $(D_x,D_y,J_z)$ as pseudospins approximately.

\section{Fermi surfaces in ordered states}
Here we discuss the FS topology in some possible multipole ordered state.  The FS was obtained by applying a finite effective field of the corresponding multipole.  Figure S5{\bf a} depicts the FS in the AFM order ($A_2^-$).  We set $0.024\Psi^1_{A_2^-}$ as the effective field.  In this case, as shown in the left figure of Fig.~S6{\bf a}, the opening gap is of the order of $4$meV, considering the renormalization of $1/10$.  The FS is similar to that in the previous DFT calculations \cite{rf:Oppeneer,rf:Elgazzar}, except for the cage around $\Gamma$, which vanishes for larger effective field.  Figure S5{\bf b} represents the FS in the $E^-$ dotriacontapole state, which is the same one as that in Fig.~4{\bf a}.  We set $0.024\bigl(\Psi^2_{E^-}+\Psi^{2'}_{E^-}\bigr)/\sqrt{2}$ as the effective field, corresponding to the experimentally observed `nematicity' along $[110]$.  To clarify differences in the FS topology, we use the same magnitude of effective field as the above $A_2^-$ AFM order.  The FS in the $E^-$ state is similar to that in the $A_2^-$ dipole state except for small separation of two FSs around $M$ and the in-plane four-fold symmetry breaking of the cage FS, which is remarkable in the comparison of $\Sigma$ line (blue line) and $\Sigma'$ line (green line) in the left figure of Fig.~S6{\bf b}.  Figure S5{\bf c} is the FS for the $A_1^-$ state with $0.024\Psi^3_{A_1^-}$.  The FS around $M$ is almost equivalent to the FS around $X$ in the paramagnetic state.  Finally, in Fig.~S5{\bf d}, we show the FS in the $E^+$ state with $0.024\bigl(\Psi^2_{E^+}+\Psi^{2'}_{E^+}\bigr)/\sqrt{2}$.  In this case, the separation between two FSs around $M$ becomes larger.  Thus, the sizable change in different ordered states may be observed in detailed analysis of the FS topology.

\begin{figure*}[h]
\includegraphics[width=0.7\textwidth]{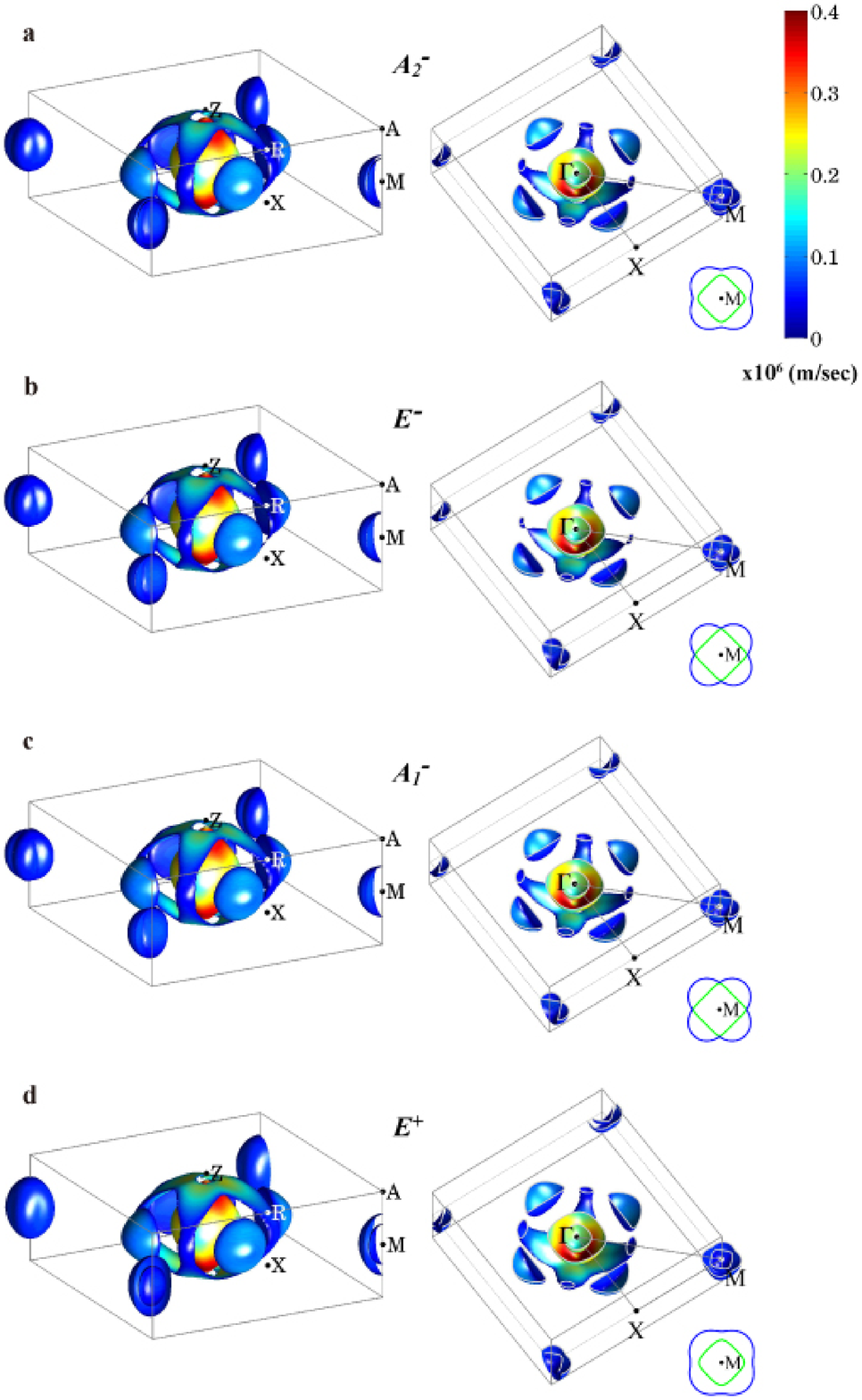}
\caption{Fermi surfaces in ordered states, colored by the Fermi velocity. The right figures are two-dimensional cut for $k_z=0$.  $A_2^-$ dipole ({\bf a}), $E^-$ dotriacontapole ({\bf b}),  $A_1^-$ dotriacontapole ({\bf c}), and $E^+$ hexadecapole ({\bf d}). The fourfold symmetry breaking in $E^-$ and $E^+$ states is verified from anisotropy of the cage FSs in ({\bf b}) and ({\bf d}). All FSs have only slight differences. The most remarkable is a change of separation between two electron FSs around $M$.}
\end{figure*}

\begin{figure*}[h]
\includegraphics[width=0.7\textwidth]{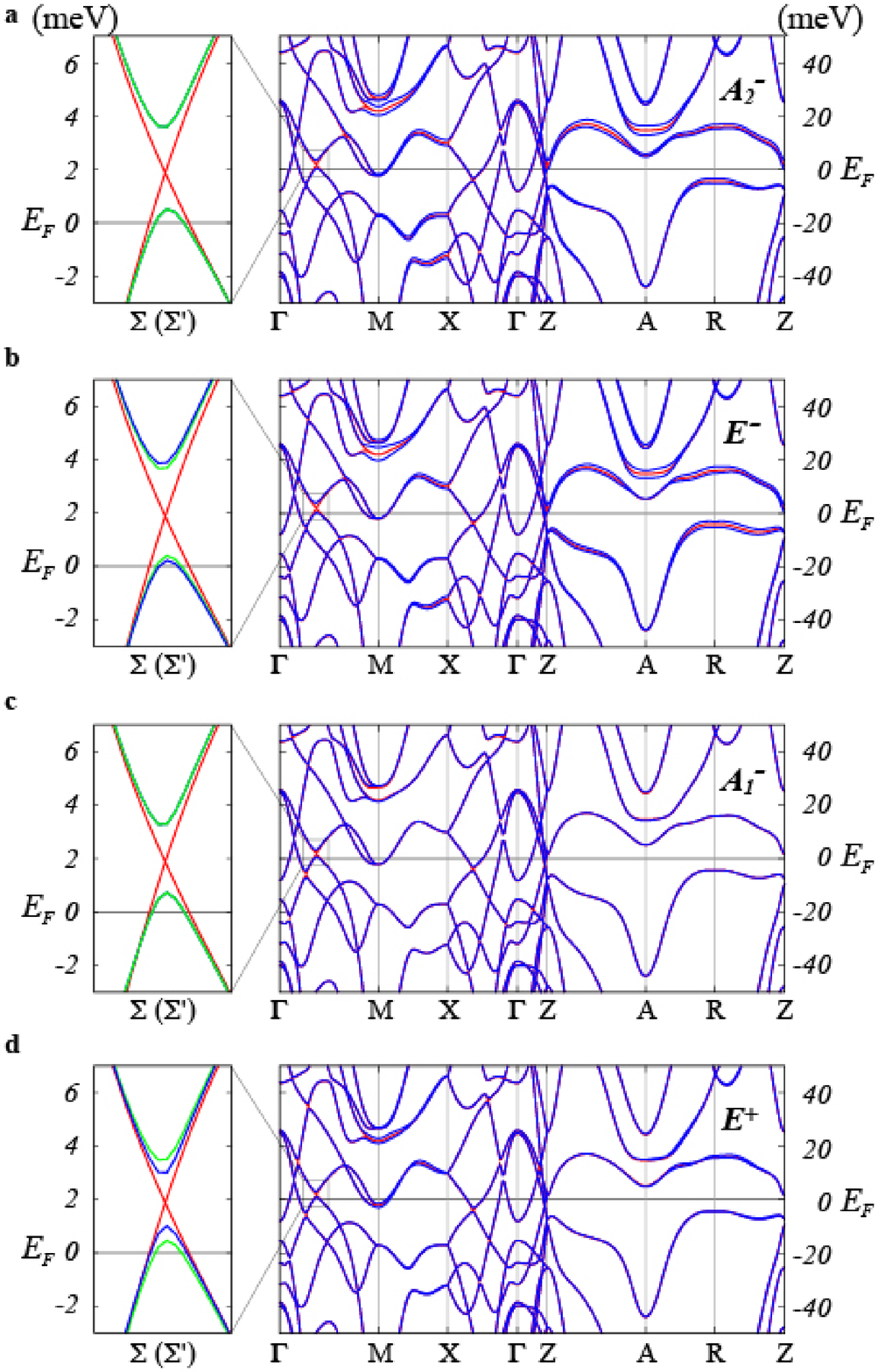}
\caption{Dispersion relations along high-symmetry line in ordered states corresponding to Fig.~S5. Red (blue) line represents dispersion relation in the paramagnetic (ordered) state. The energy scale is reduced by a factor of 10 as compared with Fig.~S1{\bf a}, by taking into account the mass renormalization effect.  Left figures are the enlarged figures around the band crossing along $\Gamma-M$ line. A large gap ($\sim 4$meV) opens as compared with the paramagnetic band (red). Blue (green) line is a dispersion relation along $\Sigma$ ($\Sigma'$) line.  In the $E^\pm$ states,  $\Sigma$ line and $\Sigma'$ line are not equivalent.}
\end{figure*}

\end{document}